\documentclass[a4paper,11pt]{article}
\usepackage{pos}

\def\be{\begin{equation}}
\def\ee{\end{equation}}
\def\beq{\begin{eqnarray}}
\def\eeq{\end{eqnarray}}
\newcommand{\bea}{\begin{eqnarray}}
\newcommand{\eea}{\end{eqnarray}}
\newcommand{\beas}{\begin{eqnarray*}}
\newcommand{\eeas}{\end{eqnarray*}}

\title{Gradient-flow renormalon subtraction and the hadronic 
tau decay series}
\author*[a]{Martin Beneke}
\author[b]{Hiromasa Takaura}

\affiliation[a]{Technische Universit\"at M\"unchen,\\
Physik Department T31,
James-Franck-Stra\ss{}e~1, 
D--85748 Garching, Germany}

\affiliation[b]{Yukawa Institute for Theoretical Physics,\\ 
Kyoto University,
Kitashirakawa Oiwakecho, Sakyo-ku, Kyoto 606-8502, Japan}

\abstract{The inconsistency between the fixed-order (FO)
and contour-improved (CI) representation of the QCD corrections 
to the inclusive hadronic tau decay width limits the precision 
to which the strong coupling can be determined from this 
process. It has recently been shown that subtracting 
the infrared renormalon divergence related to the gluon 
condensate resolves the discrepancy. Here we suggest to 
employ the gradient flow to define gauge-invariant regularized  
operators and to use the corresponding condensates in the 
operator product expansion. The associated rearrangement 
of the perturbative series results in automatic renormalon 
subtraction without the need to determine explicitly the 
Stokes constants that normalize the divergent asymptotic 
series. Applying this method to the gluon condensate, we 
find that the CI series is modified and now agrees with 
the (unmodified) FO series. This conclusively demonstrates 
the preference for the fixed-order approach, 
as has been advocated long ago.}

\FullConference{16th International Symposium on Radiative 
Corrections: Applications of Quantum Field Theory to 
Phenomenology (RADCOR2023)\\
28th May - 2nd June, 2023\\
Crieff, Scotland, UK\\[0.2cm]
{\sf TUM-HEP-1473/23, YITP-23-116, 
September 18, 2023
}}

\begin{document}
\maketitle

\section{Introduction}

\noindent
The inclusive hadronic decay width of the $\tau$-lepton 
supplies an important measurement of the strong coupling 
$\alpha_s(\mu)$ (in the $\overline{\rm MS}$ scheme) 
at the relatively low scale $\mu=m_\tau=1.777\,$GeV. 
It can be computed systematically in perturbation theory 
in terms of the operator product expansion (OPE) of the 
Adler function  $D(Q^2)$ \cite{Braaten:1991qm}. 
The perturbative coefficients are known to the five-loop 
$\mathcal{O}(\alpha_s^4)$ order 
\cite{Baikov:2008jh} and the vacuum condensate 
and other power-suppressed corrections amount 
to less than 5\% of the perturbative correction.
Making use of analyticity of the Adler function, the 
inclusive hadronic $\tau$ decay width is 
computed from a weighted integral of the Adler function 
over a circle in the complex $Q^2$-plane. 
The result can be expanded in $\alpha_s(m_\tau)$, 
a representation known as fixed-order perturbation 
theory (FOPT). Alternatively, the contour integral 
of $\alpha_s(Q^2)^n$ in the Adler function expansion is 
computed exactly without re-expansion in $\alpha_s(m_\tau)$. 
At first sight, this contour-improved 
representation 
(CIPT) \cite{Pivovarov:1991rh,LeDiberder:1992jjr} 
appears to be the method of choice, since it sums 
potentially large ``$\pi^2$-terms'' from the analytic 
continuation of logarithms into the complex plane.

However, the authors of Ref.~\cite{Beneke:2008ad} 
noted that contrary to expectation, the numerical 
difference between the FO and CI approximations does not 
decrease as one adds successively more orders, and 
traced the origin of the problem to the fact that 
QCD perturbative expansions are only asymptotic series, 
specifically due the so-called renormalon divergence
\cite{Beneke:1998ui,Beneke:2000kc}. Based on a 
plausible ansatz for the Borel transform of the 
series expansion, it was shown that the FOPT series 
approaches the Borel sum within its ambiguity, 
while CIPT does not, with the conclusion that 
CIPT should be abandoned. (A mathematical analysis 
that explains this behaviour of CIPT has appeared 
shortly before this talk  \cite{Gracia:2023qdy}.) 
Further analysis \cite{Beneke:2012vb} showed that 
the numerical difference between FOPT and CIPT is 
caused by the infrared (IR) renormalon pole related to 
the gluon condensate in the OPE despite the fact 
that its contribution to the $\tau$ decay width is 
suppressed by the weight function that relates 
the $\tau$ width to the Adler function.

The discrepancy between the FOPT and CIPT predictions 
for the hadronic $\tau$ width has been a major 
limitation for the strong coupling determination, 
since a definitive resolution of the problem should be 
such that both methods approach asymptotically the 
same value of the (appropriately defined) Borel sum.
A solution that meets this requirement was proposed 
recently \cite{Benitez-Rathgeb:2022yqb, 
Benitez-Rathgeb:2022hfj}. The idea is to subtract  
the leading IR renormalon divergence from 
the Adler function series, very 
similar to what is routinely done for the pole mass 
renormalon, where one defines (leading) renormalon-free 
quark mass schemes (see the review 
\cite{Beneke:2021lkq}). The proposed subtraction 
scheme requires that one first determines the 
normalization (Stokes constant) of the infrared 
renormalon, which can only be done approximately 
in practice, since the Stokes constant is 
non-perturbative. The authors of 
Refs.~\cite{Benitez-Rathgeb:2022yqb, Benitez-Rathgeb:2022hfj}
demonstrated that the subtraction has little 
effect on the FOPT series (due to the above-mentioned 
suppression of the gluon condensate and the associated  
renormalon series), but brings the CIPT series 
in line with the FOPT one asymptotically, in agreement 
with the earlier interpretation \cite{Beneke:2008ad} 
of the FOPT-CIPT discrepancy.  
In this method, one then combines the subtracted series 
with the ill-defined $\overline{\rm MS}$-scheme 
gluon condensate to yield a (leading) renormalon-free 
Adler function series and an unambiguous 
gluon condensate. Obtaining a non-perturbatively defined 
gluon condensate by such a combination has been attempted 
earlier \cite{Bali:2014sja} but turned out to be difficult 
in practice.

In this proceedings, we describe a renormalon subtraction 
procedure based on the gradient flow  
\cite{Luscher:2010iy,Luscher:2011bx}
that circumvents the need to obtain the 
Stokes constant and simultaneously provides a 
non-perturbative definition of the gluon condensate 
that can be implemented with existing lattice QCD 
technology. We also report first results from our study 
\cite{fullpaper} of the FO and CI hadronic $\tau$ decay 
series in this method. 

\section{Gradient-flow renormalon subtraction 
and definition of condensates}

\noindent
The connection between the gluon condensate and the leading 
IR renormalon divergence in the $\overline{\rm MS}$ 
scheme series of the Adler function has been known for a 
long time \cite{Mueller:1984vh,Beneke:1993ee}. It is also 
well-known that if the 
OPE was implemented with an explicit momentum cut-off, the 
IR renormalons in the short-distance coefficients would 
disappear and the ambiguity in subtracting the 
power divergence of the condensates would also be 
removed \cite{Novikov:1984rf}. The reason why such procedures 
have not been employed in practice 
is that simple cut-off definitions of the 
gluon condensate are not gauge-invariant and further 
it would not be possible to compute the perturbative 
series (for the Adler function) to the five-loop order 
in the presence of a cut-off. 

We propose to define the gradient-flow regularized 
gluon condensate as the 
matrix element of the product of the corresponding fields 
at finite flow time $t$. The basic idea is general, and 
applies to other matrix elements of local 
operators appearing in the OPE. It works because the 
gradient-flowed fields are smeared fields and 
composite operators of flowed fields do not need operator 
renormalization. The power-divergences of local 
higher-dimension operators reappear as singular terms 
as $t\to 0$. Thus, gradient-flowed local operators 
are naturally defined with a cut-off or order $1/\sqrt{t}$ 
with the crucial advantage that gauge-invariance 
is preserved. The gradient-flowed gluon condensate 
is identical to the so-called action density
\begin{equation}
E(t) 
%\equiv \pi^2 \langle \widetilde{\frac{\alpha_s}{\pi} G^2} \rangle(t)
\equiv \frac{g_s^2}{4} \langle 0|\widetilde{G}^A_{\mu\nu}(t) 
\widetilde{G}^{A\mu\nu}(t)|0\rangle\,, 
\end{equation}
where the matrix-valued field strength is defined in terms of the 
flowed gluon field \cite{Luscher:2010iy}
\begin{equation}
\widetilde{G}_{\mu\nu}(t) = 
\partial_\mu B_\nu(t) - \partial_\nu B_\mu(t) 
-i g_s \left[B_\mu(t),B_\nu(t)\right]
\end{equation}
by the standard expression. 
The action density has been studied non-perturbatively on 
the lattice as a function of the flow parameter $t$ 
\cite{Luscher:2010iy}. For small flow time, 
its OPE is given by
\begin{equation}
E(t) = \pi^2 \left(
\frac{\widetilde{C}_1(t)}{t^2} + \widetilde{C}_{G^2}(t) 
\langle \frac{\alpha_s}{\pi}G^2\rangle+\mathcal{O}(t)\right).
\label{eq:EOPE}
\end{equation}
The most singular term $1/t^2$ corresponds to the quartic power  
divergence of the local operator $G^2=G_{\mu\nu}^A G^{A\mu\nu}$ 
and its mixing into the unit operator with coefficient 
$\widetilde{C}_1(t)$, which can be computed perturbatively 
in the strong coupling. The next term involves the 
usual local gluon condensate, which we define as the 
scale-invariant gluon condensate
\begin{equation}
\langle \frac{\alpha_s}{\pi} G^2\rangle 
\equiv \frac{\beta(\alpha_s)}{\pi\beta_0 \alpha_s} 
\langle 0| G^A_{\mu\nu} G^{A\mu\nu}|0\rangle\,.
\end{equation}
The coefficient $\widetilde{C}_1(t)$ has been computed 
to $\mathcal{O}(\alpha_s^2)$ in \cite{Luscher:2010iy} and 
$\mathcal{O}(\alpha_s^3)$ in \cite{Harlander:2016vzb}, 
the coefficient $\widetilde{C}_{G^2}(t)$ is known to 
next-to-next-to-leading order (NNLO) $\mathcal{O}(\alpha_s^2)$ 
from \cite{Harlander:2018zpi}.

\section{Gradient-flow renormalon-subtracted 
Adler function}
\label{sec:Adlersub}

\noindent 
The Adler function is defined in terms of the 
vector-current two-point function, and expanded as 
\begin{eqnarray}
D(s) &\!\!\!=\!\!\!& -s\frac{d \Pi(s)}{ds} =  
\frac{N_c}{12\pi^2} \sum\limits_{n=0}^\infty a_\mu^n
\sum\limits_{k=1}^{n+1} k\, c_{n,k}\,\ln^{k-1}\frac{-s}{\mu^2} 
= \frac{N_c}{12\pi^2} \sum\limits_{n=0}^\infty c_{n,1}\,a_Q^n 
\nonumber \\[0.2cm]
&\!\!\!=\!\!\!& \frac{N_c}{12\pi^2} 
\left[1 + a_Q + 1.64 a_Q^2 + 6.37 a_Q^3 + 49.08 a_Q^4 + 
\ldots\right].
\label{eq:adlerpert}
\end{eqnarray}
Here $s=-Q^2$ and $a_\mu=\alpha_s(\mu)/\pi$ ($c_{n,n+1}=0$ 
except for $n=0$). The series 
is known up to the five-loop order \cite{Baikov:2008jh}. The 
numerical values refer to the relevant case with $n_f=3$ massless 
quark flavours. Including the leading gluon condensate 
correction in the OPE, we may write
\begin{equation}
D(Q^2)  = \frac{N_c}{12\pi^2}\left(C_1(Q^2)+
\frac{C_{G^2}(Q^2)}{Q^4} \,  \langle \frac{\alpha_s}{\pi} G^2\rangle  
+\mathcal{O}(1/Q^6)\right).
\label{eq:adlerOPE}
\end{equation}
The coefficient function $C_G^2(Q^2)$ of the gluon operator 
is known to $\mathcal{O}(\alpha_s)$ 
\cite{Chetyrkin:1985kn} and NNLO 
$\mathcal{O}(\alpha_s^2)$ \cite{Harlander1998}.
The expansion of $C_1(Q^2)$ given by the second line of 
\eqref{eq:adlerpert} is an asymptotic series. The leading 
divergence arises from an ultraviolet (UV) renormalon. Its normalization 
(Stokes constant) turns out to be numerically small. At 
intermediate perturbative orders, the dominant component 
of the asymptotic expansion is of IR origin. The corresponding 
IR renormalon series takes the form
\begin{eqnarray} 
C_1^{\rm IR}(Q^2) = 
C_{G^2}(Q^2)\,\frac{\mu^4}{Q^4}\,
\sum_n \alpha_s^{n+1}(\mu) K \left(-\frac{\beta_0}{a}\right)^{\!n} 
n! \,n^b \left(1+\frac{s_1}{n}+O(1/n^2)\right),
\label{eq:IR2}
\end{eqnarray}
where $a=2$ is related to the dimension of the gluon 
condensate operator, $b, s_1$ to 
the QCD beta function, and $K$ is the unknown Stokes constant. 
The relation is a consequence of the fact that the 
Adler function is an observable, hence the ambiguity 
in defining the sum of the asymptotic series must be 
fixed by whatever scheme one chooses to define the 
gluon condensate non-perturbatively.

Since the action density provides such a definition, 
the IR renormalon divergence of $\widetilde{C}_1(t)$ 
must have exactly the same form as \eqref{eq:IR2} with 
$C_{G^2}(Q^2)$ replaced by $\widetilde{C}_{G^2}(Q^2)$. We 
therefore solve \eqref{eq:EOPE} for 
$\langle \frac{\alpha_s}{\pi}G^2\rangle$, and eliminate it from 
\eqref{eq:adlerOPE} to obtain
\begin{equation}
D(Q^2)  = \frac{N_c}{12\pi^2}\Bigg(\,
\underbrace{\left[C_1(Q^2)-
\frac{r}{t^2 Q^4} \,\widetilde{C}_1(t)\right]}_{\mbox{
\footnotesize renormalon cancels}}+
\underbrace{\frac{r}{Q^4} \, \frac{E(t)}{\pi^2}}_{\mbox{
\footnotesize nonpert. defined}}
+\;\mathcal{O}(1/Q^6)\,\Bigg)
\label{eq:adlerOPEsub}
\end{equation}
with 
\begin{equation}
r = \frac{C_{G^2}(Q^2)}{\widetilde{C}_{G^2}(t)} 
= \frac{2\pi^2}{3}\left(\frac{1}{6} - 
\frac{35}{24}\,\frac{\alpha_s(\mu)}{\pi} 
+\mathcal{O}(\alpha_s^2)\right).
\label{eq:r}
\end{equation}
The key advantage of \eqref{eq:adlerOPEsub} over 
\eqref{eq:adlerOPE} is that the subtraction solves 
both problems of the standard OPE: the perturbative series in 
square brackets is free of the leading IR renormalon 
divergence, while the leading non-perturbative correction  
is now well-defined and can be computed directly on the 
lattice. 

In the following, we discuss the perturbative expansion of 
the subtracted Adler function. Both, the Adler function and 
the subtraction term are known to rather high orders from 
the perspective of multi-loop computations, but there is a 
slight mismatch as the subtraction term is available only 
to $\mathcal{O}(\alpha_s^3)$. In previous studies of the 
behaviour of the perturbative expansion \cite{Beneke:2008ad}, 
it proved instructive 
to merge the exactly known low-order coefficients with the 
asymptotic behaviour to model the 
series expansion to all orders. For the Adler 
function, an estimate of the $\mathcal{O}(\alpha_s^5)$ 
coefficient is included, and then $c_1, \ldots c_5$ 
are employed to determine the five unknown parameters 
of the ansatz 
\begin{equation}
B[D](u) \,=\, B[D_1^{\rm UV}](u) + 
B[D_2^{\rm IR}](u) + B[D_3^{\rm IR}](u) +
d_0^{\rm PO} + d_1^{\rm PO} u 
\end{equation}
for the Borel transform of the Adler function. The first 
three terms on the right-hand side incorporate the first 
UV renormalon and first two IR renormalon singularities with unknown 
Stokes constants, which together with $d_0^{\rm PO}$, $d_1^{\rm PO}$
are chosen such that  $c_1, \ldots c_5$ are reproduced 
exactly from the expansion of the Borel transform in 
$u$. We refer to \cite{Beneke:2008ad} for the explicit 
expressions and details. 

For the subtraction term, three low-order terms 
$e_1, e_2, e_3$ in the expansion of $E$ are available. 
However, the series expansion of $E$ has no UV renormalons, 
while the Stokes constant of the gluon condensate 
renormalon series $E_2^{\rm IR}$ is tied to  
$D_2^{\rm IR}$ to effect the renormalon cancellation 
related to the universal gluon condensate. Hence no 
further information is required to determine the 
three parameters of the ansatz
\begin{equation}
B[E](u) \,=\, B[E_2^{\rm IR}](u) + B[E_3^{\rm IR}](u) +
e_0^{\rm PO} + e_1^{\rm PO} u 
\end{equation}
for the Borel transform of the series expansion of the 
subtraction term in terms of the three exactly known 
coefficients $e_1, e_2, e_3$.

For the following analysis we implement the above-described 
Borel transform model of 
\cite{Beneke:2008ad}, generalized to 5-loop accuracy for 
the QCD beta-function and including the $\mathcal{O}(\alpha_s^2)$ 
contribution to $C_{GG}(Q^2)$ \cite{Harlander1998}.  
The expression for $r$ is included in the present preliminary 
analysis only to $\mathcal{O}(\alpha_s)$. 
We set $\mu=m_\tau$, $\alpha_s(m_\tau)=0.34$, 
and the gradient-flow time to 
$8 t=20/m_\tau^2$, which corresponds to a low UV cut-off 
on the gradient-flow regularized gluon condensate at the 
limit of perturbativity. 

\begin{figure}
\centering
\includegraphics[width=0.71\linewidth]{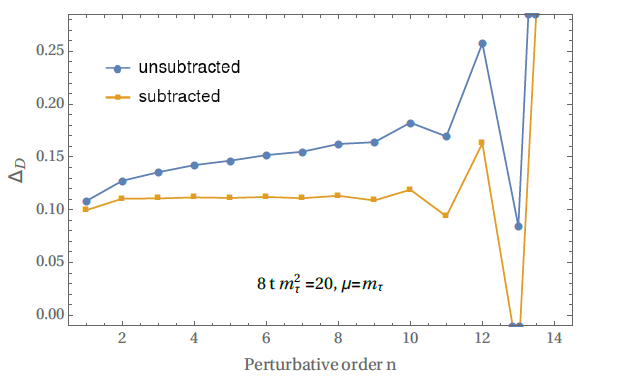}
\caption{Unsubtracted and gradient-flow subtracted Adler-function 
series, summed to perturbative order $n$.}
\label{fig:adler}
\end{figure}

In Figure~\ref{fig:adler} we show the cumulative partial 
sum to order $n$ of the series expansion of the 
perturbative correction 
to the Adler function, more precisely to $\Delta_D$, 
defined as
\begin{equation}
D(s) = \frac{N_c}{12\pi^2}\left[1+\Delta_D\right].
\end{equation}
The upper (blue) points show the unsubtracted Adler function, 
which increases slowly with order until after the 10th 
order the sign of the added terms begins to alternate 
as a consequence of the dominance of the first UV 
renormalon. The sign alternation becomes visible only at 
such high orders as the normalization of UV renormalons 
is suppressed in the $\overline{\rm MS}$ scheme 
\cite{Beneke:1992ea} as can be checked explicitly in the 
so-called large-$\beta_0$ approximation \cite{Beneke:1992ch, 
Broadhurst:1992si, Ball:1995ni}. The renormalon-subtracted Adler 
function (lower, orange points) hares this feature, as the subtraction term 
does not affect the UV renormalon behaviour. We then 
observe that the main difference to the standard unsubtracted 
series \eqref{eq:adlerpert} is that 
in intermediate orders the subtracted series terms 
are very small and the partial sum up to the 
8th order almost coincides with the subtracted Adler functions 
at $\mathcal{O}(\alpha_s^2)$. 

The Adler function including condensate 
corrections should be independent of the subtraction 
term, which suggests that the gradient-flow regularized 
gluon condensate differs significantly from the standard one. 
The standard definition is ambiguous, but comparison 
with Figure~6 of \cite{Beneke:2008ad} shows that the 
ambiguity is smaller than the difference between the 
two sets of points in Figure~\ref{fig:adler} 
near the minimal term of the unsubtracted 
series at the fifth perturbative order. Nevertheless, for 
the following analysis of the hadronic $\tau$ lepton width 
in the FO and CI expansion, the value of the gradient-flow 
regularized gluon condensate is not important, since 
its contribution is suppressed by two powers of 
$\alpha_s$ relative to the Adler function.

%%%%%%%%%%%%%%%%%%%%%%%%%%%%%%%%%%%%%%%%
%%%%%%%%%%%%%%%%%%%%%%%%%%%%%%%%%%%%%%%%%%%%%%

\section{Hadronic tau decay series}
\label{sec:taudecay}

\noindent 
Turning to the hadronic $\tau$ lepton width, we write the QCD 
contribution in the standard form
\begin{equation}
R_\tau = -\,i\pi\!\oint\limits_{|x|=1} \frac{dx}{x}\,(1-x)^3
\, 3\,(1+x)\,2D(M_{\tau}^2 x)
= N_c \,\biggl[\,
1 + \delta^{(0)} +\mbox{power corrections}\biggr]\,.
\end{equation}
The perturbative correction to the parton-model 
prediction $R_\tau=3$ is quantified by $\delta^{(0)}$. 
From \eqref{eq:adlerpert} one finds 
\begin{equation}
\delta^{(0)}_{\rm FO} = \sum\limits_{n=1}^\infty a(M_\tau^2)^n
\sum\limits_{k=1}^{n} k\,c_{n,k}\,J_{k-1}
\quad\mbox{with}
\quad J_l \equiv \frac{1}{2\pi i} \!\!\oint\limits_{|x|=1} \!\!
\frac{dx}{x}\, (1-x)^3\,(1+x) \ln^l(-x)
\label{eq:FOser}
\end{equation}
in FOPT, and 
\begin{equation}
\delta^{(0)}_{\rm CI} = \sum\limits_{n=1}^\infty c_{n,1}\,
J_n^a(M_\tau^2)
\quad\mbox{with} 
\quad J_n^a(M_\tau^2) \equiv \frac{1}{2\pi i} \!\!\oint\limits_{|x|=1}\!\!
\frac{dx}{x}\,(1-x)^3\,(1+x)\,a^n(-M_\tau^2 x) 
\label{eq:CIser}
\end{equation}
in CIPT, resulting in the following expansions 
for the first four exactly known (in brackets: plus the 
estimate of the fifth) terms:
\begin{eqnarray}
\label{del0FOn}
\delta^{(0)}_{\rm FO} &=&
0.1082 + 0.0609 + 0.0334 + 0.0174\,(+\,0.0088\,) \,=\, 0.2200 \;(0.2288) 
\\[0.1cm]
%\tvs
\label{del0CIn}
\delta^{(0)}_{\rm CI} &=&
0.1479 + 0.0297 + 0.0122 + 0.0086\,(+\,0.0038\,) \,=\, 0.1984 \;(0.2021) 
\end{eqnarray}
The observation that the partial sums do not approach each 
other upon adding successive terms 
illustrates the FOPT-CIPT discrepancy.

\begin{figure}[t]
\centering
\includegraphics[width=0.7\linewidth]{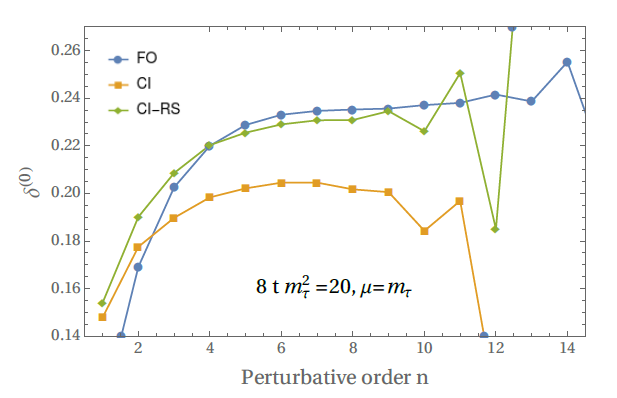}
\caption{Perturbative correction $\delta^{(0)}$ to the inclusive hadronic 
$\tau$ lepton width summed to order $\alpha_s^n$ in FOPT (blue), 
CIPT (orange) and renormalon-subtracted CIPT (green).}
\label{fig:tau}
\end{figure}

We extend~\eqref{eq:FOser}, \eqref{eq:CIser} to include the 
subtraction term of the Adler function. It is important to 
note that the subtraction term contributes to the FO 
coefficient only when it contains a $\ln Q^2/\mu^2$, which 
can appear only through $C_{G^2}$ in $r$. As can be seen from 
\eqref{eq:r}, this happens only from $\mathcal{O}(\alpha_s^2)$. 
Hence, to the NLO $r$-order employed here, the unsubtracted 
and gradient-flow renormalon-subtracted FO series 
for the hadronic $\tau$ decay width are identical. This 
reflects the mentioned suppression of the gluon condensate 
contribution to inclusive tau decay. As a consequence, the power 
correction stemming from the gluon condensate is a sub-percent 
effect numerically. To the contrary, the gradient-flow 
subtraction contributes to the CI series from the first order 
in $\alpha_s$, which demonstrates explicitly that the CI 
prescription is incompatible with the OPE. Numerically, 
the renormalon-subtracted CI series reads in the first 
five orders:
\begin{eqnarray}
\label{del0CInsub}
\delta^{(0)}_{\rm CI,RS} &=&
0.1542 + 0.0362 + 0.0185 + 0.0118\,(+\,0.0053\,) \,=\, 0.2207 \;(0.2259) 
\end{eqnarray}
Comparing this to the FO series \eqref{del0FOn}, we observe that 
the two series are now close to each 
other from the fourth order. Figure~\ref{fig:tau} shows 
the FO, CI and subtracted CI-RS series in higher 
orders adopting the Borel function ansatz discussed above. 
This demonstrates conclusively that the renormalon-subtraction 
via gradient flow solves the original FOPT-CIPT discrepancy 
in favour of FOPT as advocated in \cite{Beneke:2008ad}

%%%%%%%%%%%%%%%%%%%%%%%%%%%%%%%%%%%%%%%%
%%%%%%%%%%%%%%%%%%%%%%%%%%%%%%%%%%%%%%%%%%%%%%

\section{Conclusion}
\label{sec:summary}

\noindent 
The discrepancy between the fixed-order 
and contour-improved computation of the 
perturbative QCD corrections to the inclusive hadronic 
$\tau$-lepton decay width \cite{Beneke:2008ad, Beneke:2012vb} 
has limited the precision 
to which the strong coupling can be determined from this 
process. The issue was recently resolved by gluon-condensate 
renormalon subtraction \cite{Benitez-Rathgeb:2022yqb, 
Benitez-Rathgeb:2022hfj}. In the present work we 
suggested a new method to perform renormalon subtraction 
by defining the local operators in the OPE by 
their gradient-flow representation at finite flow time $t$. 
The flow time acts as a gauge-invariant UV cut-off.  
The rearrangement of perturbative corrections from 
the $\overline{\rm MS}$ perturbative series into the 
gradient-flow regularized operator automatically subtracts 
the IR renormalon divergence of the series associated 
with the corresponding operator. The advantage of this 
method over previous ones is that it avoids the determination 
of the normalization (Stokes) constant of the renormalon 
series while simultaneously providing a non-perturbatively 
valid definition of the condensates, allowing for a 
consistent addition of power corrections to the perturbative 
series. The flowed operators at the required value of 
$t$ can be computed 
on the lattice. The gradient flow separates the continuum 
limit $a\to 0$ on the lattice from the cut-off scale $1/\sqrt{t}$ 
defining the renormalon subtraction, and allows one to define 
cut-off condensates in the continuum.

Applying this method to the gluon condensate, we 
find that the CI series for the hadronic tau width 
is significantly modified despite the fact that the gluon 
condensate makes a numerically negligible correction 
to the OPE for inclusive $\tau$-lepton decay. 
Within the approximations 
employed in the present study, the FO series remains 
unaltered by the subtraction due to the inclusive spectral 
weight function that suppresses the gluon condensate by 
two powers of $\alpha_s$. Fig.~\ref{fig:tau} shows 
the three series expansions (FO, CI, CI-renormalon subtracted)
and conclusively demonstrates 
the preference for the FO approach (as advocated 
in \cite{Beneke:2008ad}), since the FO 
and subtracted CI series are in full agreement. 

The present analysis should be refined by considering 
the dependence of the subtraction on the renormalization 
scale $\mu$ and flow time $t$. In principle, $r$ from
\eqref{eq:r} is already available to $\mathcal{O}(\alpha_s^2)$. 
Including this term causes a technical complication, 
since the subtraction term acquires a logarithmic 
$Q$-dependence at this order, but has little numerical effect on the 
final result. The details of this analysis will be 
presented in \cite{fullpaper}. 

%\cite{Ball:1995ni} Large-b0 tau with Ball-Braun\\
%\cite{Beneke:1992ch} My Adler fn Borel trafo 1992\\
%\cite{Beneke:2016cbu} top mass renormalon residue\\
%\cite{Broadhurst:1992si} Broadhurst Adler large-b0\\
%\cite{Iritani:2018idk} Hiromasa I\\
%\cite{Suzuki:2021tlr} Hiromasa II\\

\bibliographystyle{JHEP} % bst file
\bibliography{refs.bib}

%\begin{thebibliography}{99}
%\bibitem{...}
%....
%\end{thebibliography}

\end{document}